 \documentclass[final,5p,times,twocolumn]{elsarticle}

 \usepackage{graphicx}

\usepackage{amssymb}
\usepackage{amsmath}
\journal{Physica C}

\newcommand{\co}        {$^{59}$Co}
\newcommand{\ind}        {$^{115}$In}

\newcommand{\cecoin}    {CeCoIn$_5$}

\begin{document}

\begin{frontmatter}



\title{NMR Studies of Field Induced Magnetism in CeCoIn$_5$}


\author[UCD]{Nicholas J. Curro}
\author[Taiwan]{Ben-Li Young}
\author[NHMFL]{Ricardo R. Urbano}
\author[LANL]{Matthias J. Graf}
\address[UCD]{Department of Physics, University of California, Davis, CA 9516, USA}
\address[Taiwan]{Department of Electrophysics, National Chiao Tung University, Hsinchu 30010, Taiwan}
\address[NHMFL]{National High Magnetic Field Laboratory, Florida State University, Tallahassee, Florida 32306-4005, USA}
\address[LANL]{Theoretical Division, Los Alamos National Laboratory, Los Alamos, New Mexico 87545, USA}

\begin{abstract}
Recent Nuclear Magnetic Resonance and elastic neutron scattering experiments have revealed conclusively the presence of static incommensurate magnetism in the field-induced B phase of \cecoin.  We analyze the NMR data assuming the hyperfine coupling to the In(2) nuclei is anisotropic and simulate the spectra for several different magnetic structures.  The NMR data are consistent with ordered Ce moments along the [001] direction, but are relatively insensitive to the direction of the incommensurate wavevector.
\end{abstract}


\end{frontmatter}


\section{Introduction}

The heavy fermion superconductor \cecoin\ exhibits a rich spectrum of strongly correlated electron behavior.  This unconventional d-wave superconductor exhibits non-Fermi liquid behavior associated with proximity to a proposed quantum critical point, as well as a new thermodynamic phase
(B phase) that exists only within the superconducting phase near $H_{c2}$ \cite{sidorov,romanQCPCeCoIn5}.
Initially this B phase was identified as the  elusive Fulde-Ferrell-Larkin-Ovchinnikov (FFLO) superconducting phase first predicted to exist
in Pauli-limited superconductors over 40 years ago \cite{andrea,romanFFLO,radovanCeCoIn5FFLO,kumagaiCeCoIn5}.
However, recent NMR work identified the presence of incommensurate antiferromagnetic order in the B phase in contrast to the standard predictions for the FFLO
phase \cite{CurroCeCoIn5FFLO,graf,grafknight}.
Signatures of magnetism were also seen in other NMR experiments \cite{mitrovic, mitrovicKvsHCo115}.
Despite initial arguments to the contrary \cite{MatsudaFFLOReview},
 recent neutron scattering results by Kenzelmann and coworkers now provide conclusive proof for long-range static incommensurate antiferromagnetic order \cite{KenzelmannCeCoIn5Qphase}.

The original NMR work measured the spectrum of the In(2) sites and proposed a candidate magnetic structure in which both the ordered Ce moments  and the incommensurate wavevector are parallel to the applied field (along [100]).  However, the neutron scattering experiments found that when the field was applied along [$1\bar{1}0$] the moments lie along [001] and the incommensuration along [110].  Recently we have re-analyzed the NMR data and shown that by allowing for \textit{anisotropic} hyperfine coupling between the Ce spins and the \ind\ nuclei, rather than simple isotropic coupling as originally assumed in \cite{CurroCeCoIn5FFLO}, then the NMR spectra are fully consistent with moments along [001]  \cite{CurroJLTP2009}.  Here, we show that by using hyperfine parameters consistent with experimental measurements of the Knight shift, we find quantitative agreement for the \ind\ NMR spectra and put constraints on the incommensuration wavevector.

\section{Incommensurate Antiferromagnetism}

\begin{figure}
\begin{center}
\includegraphics[width=0.25\textwidth]{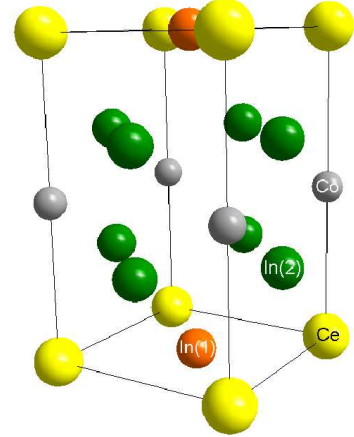}
\end{center}
\caption{(Color online) The unit cell of \cecoin.  The Ce atoms (yellow) sit at the eight corners.
The In(1) atoms sit in the center of the top and bottom faces (orange).
The Co atoms are grey and the In(2) atoms are green.
For the field oriented in the $ab$ plane, there are two inequivalent In(2) atoms,
depending on whether the field is parallel, In(2a), or perpendicular, In(2b), to the unit cell face.}
\label{fig:unitcell}
\end{figure}

The incommensurate magnetic structure is given by $\mathbf{S}(\mathbf{r}) = \mathbf{S}_0 \cos[(\mathbf{Q}_0+\mathbf{Q}_i)\cdot\mathbf{r}]$, where $\mathbf{S}_0$ is the ordered moment,
with commensurate,
$\mathbf{Q}_0 = (\pi/a,\pi/a,\pi/c)$,
and incommensurate,
$\mathbf{Q}_i = \pi/a(\delta_x,\delta_y,0)$,
antiferromagnetic wavevectors.
Kenzelmann et al. found $\delta_x=\delta_y=0.12$ and $S_0=0.15\mu_B$ along [001].  The ordered moments create a static field at the \ind\ and \co\
nuclear sites through the hyperfine coupling. This field cancels by symmetry at the In(1) and Co sites, but the NMR spectra show a large hyperfine field at the In(2a) site that is aligned either parallel or antiparallel to the applied field (along [100]).  The hyperfine coupling to the In(2a) site  is given by $\mathcal{H} = \sum_{i\in nn}\hat{\mathbf{I}}\cdot\mathbb{B}\cdot\mathbf{S}(\mathbf{R}_i)$, where the sum is over the two nearest neighbor Ce spins (along [100]). The coupling tensor is given by $\mathbb{B} =B_{\rm iso}\mathbb{I}+\mathbb{B}_{\rm dip}$,
\begin{align}
\label{eqn:hypcoupling}
    &\mathbb{B}_{dip} =\\
\nonumber &B_{\rm dip}\left(
\begin{array}{ccc}
 \frac{1}{2}(1-3\cos 2\theta_z) & 0 & \pm\frac{3}{2}\sin 2 \theta_z \\
 0 & -1 & 0 \\
 \pm\frac{3}{2}\sin 2 \theta_z & 0 & \frac{1}{2}(1+3\cos 2\theta_z)
\end{array}
\right).
\end{align}
 The constants $B_{iso}$, $B_{dip}$ and $\theta_z$ can be determined by fits to the Knight shift and are given by: $B_{iso}=B_{dip}=7.1$ kOe/$\mu_B$ and $\theta_z=29^{\circ}$ \cite{CurroJLTP2009}.  In the B phase, the static hyperfine field at the In(2a) site varies spatially with
\begin{align}
\label{eqn:hypfield}
\nonumber \mathbf{H}_{hf}(2a) = S_0B_{dip}
\big[ &3\sin(2\theta_z)\cos(\frac{\pi\delta_x}{2})\hat{\mathbf{a}}\\
&+ \left( (1-\cos(2\theta_z))\sin(\frac{\pi\delta_x}{2}) \right) \hat{\mathbf{c}} \big].
\end{align}

\begin{figure}
\begin{center}
\includegraphics[width=0.48\textwidth]{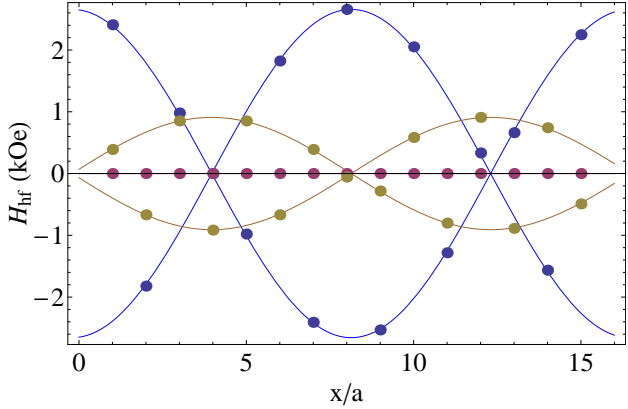}
\end{center}
\caption{(Color online) The static hyperfine field at the In(2a) site along the [100] (blue),  [010] (purple) and [001] (brown) directions as a function of lattice position along [100] for $S_0=0.15\mu_B$ and $\delta_x=\delta_y=0.121$,
with lattice parameter $a$, respectively.
The solid lines show the modulation of the staggered magnetization (Eq. \ref{eqn:hypfield}).
}
\label{fig:hypfieldmap}
\end{figure}

\begin{figure*}
\begin{center}
\includegraphics[width=1.0\textwidth]{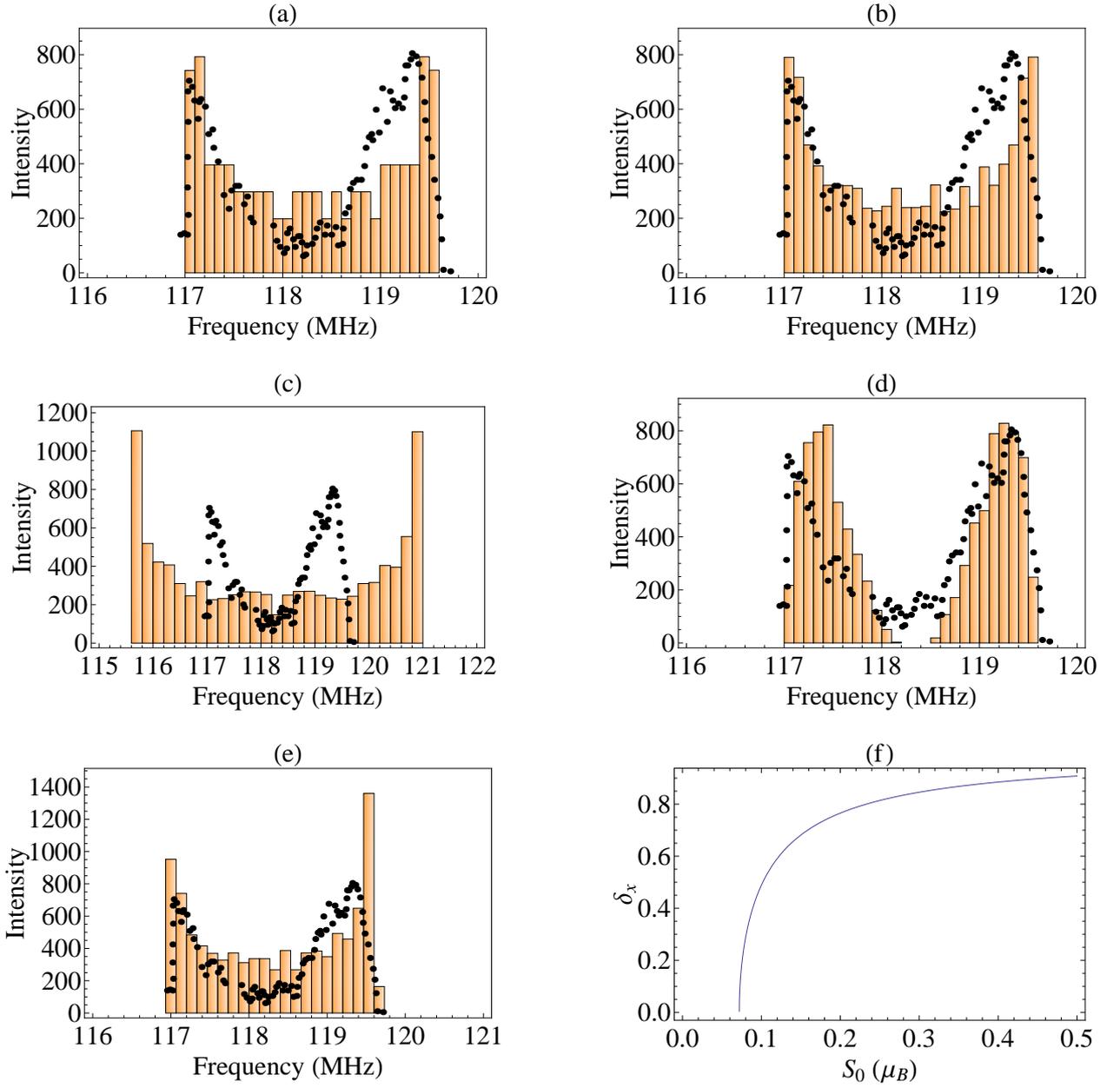}
\end{center}
\caption{(Color online) Simulated spectra and experimental data at 11.1 T and 50 mK (solid points, reproduced from \cite{CurroCeCoIn5FFLO}) for
(a) $(\delta_x,\delta_Y)=(0,0.121)$ and $S_0=0.07\mu_B$,
(b) $(\delta_x,\delta_Y)=(0.121,0.121)$ and $S_0=0.07\mu_B$,
(c) $(\delta_x,\delta_Y)=(0.121,0.121)$ and $S_0=0.15\mu_B$,
(d) $(\delta_x,\delta_Y)=(0.498,0.498)$ and $S_0=0.10\mu_B$,
(e) $(\delta_x,\delta_Y)=(0.681,0.681)$ and $S_0=0.15\mu_B$.
(f) The locus of points $(S_0, \delta_x)$ that are consistent with experiment.
}
\label{fig:spectra}
\end{figure*}

In order to make detailed comparisons with experimental spectra, we have calculated the hyperfine field and the resonance frequency, $f = \gamma|\mathbf{H}_0+\mathbf{H}_{hf}|$, where $\gamma\mathbf{H}_0 ~||~[100]\sim 118.3$ MHz, for several different values of $\delta_x$ and $\delta_y$ for a lattice of 100 x 100 unit cells. Representative cuts in real space are shown in Fig. \ref{fig:hypfieldmap}. Spectra were determined by calculating the histogram of frequencies, shown in Fig. \ref{fig:spectra}.
As seen in Fig. \ref{fig:hypfieldmap}, the hyperfine field at the In(2a) site has components either parallel or antiparallel to the  applied field leading to the broad double-peak spectra seen in Fig. \ref{fig:spectra}.  However, the size of the ordered moment measured by neutrons, $S_0=0.15\mu_B$ implies a hyperfine field larger than observed by NMR.  The NMR data are more consistent with an ordered moment of 0.07$\mu_B$.  Using Eq. \ref{eqn:hypfield} and the measured value of the modulus of the hyperfine field of 1.3 kOe, we find that the ordered moment and incommensuration are related by $\delta_x=\delta_{x0}\cos^{-1}(S_0^0/S_0)$, where $\delta_{x0} = 0.64$ and $S_0^0 = 0.07~\mu_B$ (shown in Fig. \ref{fig:spectra}(f)).

As seen in Fig. \ref{fig:spectra}(a) - \ref{fig:spectra}(e), the measured spectra is relatively insensitive to the direction of the incommensurate wavevector $\mathbf{Q}_i$.  Subtle changes in the calculated spectra arise because of the component of $\mathbf{H}_{hf}(2a)$ along [001] gives rise to small frequency shifts.

The difference between the magnitude of the ordered moment measured by neutrons versus that measured by NMR may be due to either (i) changes in the magnetic structure as a function of the orientation of the field, (ii) errors associated with the measured values of the hyperfine coupling, (iii) dynamic fluctuations of the magnetic structure,
or (iv) neglect of coupling to conduction electrons.
There are two types of hyperfine couplings in heavy-fermion compounds, one to the conduction electron spins,
$\mathbf{S}_c$, and another to the local moments, $\mathbf{S}_f$ \cite{CurroKSA}.
Here, we ignored the coupling to $\mathbf{S}_c$,
since the static hyperfine field presumably arises from the ordered Ce moments in the B phase.
Including a hyperfine coupling to the $\mathbf{S}_c$ spins may account for the difference in the ordered moments and
the incommensuration indicative of a Fermi surface instability.

In the original measurements of the In(2a) NMR, the broad spectrum only emerged below $\sim 100$ mK (at 11.1 T where $T_c=470$ mK and $T_{0}=290$ mK, where $T_0$ is the transition temperature to the B phase).  Above $\sim 100$ mK the In(2a) spectrum disappeared.  The reason for this behavior is likely due to dynamic effects, in which the spin-spin relaxation rate $T_2^{-1}$ is excessively fast due to the dynamics of the magnetic structure.  For sufficiently low temperatures where the magnetic structure becomes static and $T_2^{-1}$ is small enough the broad spectrum emerges.  However, when the time scale of the fluctuations is of the inverse of the linewidth ($\sim 2.8$ kHz in this case) the splitting of the NMR spectrum can be less than the full static linewidth.  It is possible that for temperatures $T\ll 60$ mK, the NMR linewidth would broaden to reflect a larger ordered moment.

\section{Superconductivity}

An outstanding question regarding the nature of the B phase is the possible modulation of the superconducting order parameter
$\Delta$. Using Ginzburg-Landau theory, Kenzelmann et al.\  proposed a specific coupling between $\Delta$,
the antiferromagnetic order parameter $M_\mathbf{Q}$, and the magnetic field ${\bf H}$  to account for the phase
diagram \cite{KenzelmannCeCoIn5Qphase}.
They postulated that at large fields an induced subdominant
order parameter nucleates with finite momentum, $\Delta_\mathbf{Q}$,
in addition to the dominant, uniform $d$-wave order parameter $\Delta_0$.
In real space, this scenario suggests that the superconductivity and antiferromagnetism are out of phase, such that the extrema of
magnetization $M_\mathbf{Q}$ are located at the nodes of $\Delta_\mathbf{Q}$.
Consequently, $\Delta_\mathbf{Q}$ should vary at atomic length scales,
which is physically unreasonable considering the stiffness of the condensate, $2\pi/|\mathbf{Q}| \approx 2 a \ll \xi_0$.

The NMR spectra of the In(1) site, where the hyperfine field cancels, show no broadening that would be associated with
a spatial variation of the Knight shift $K$
due to real-space nodes \cite{CurroCeCoIn5FFLO}.
$K$ is proportional to the spin susceptibility, which vanishes for $T\ll \Delta/k_B$ \cite{CurroHFreview}.  If $\Delta$ varied spatially and vanished at real-space nodes,
then $K$ would vary as well leading to a variation of the resonance frequency and a broad resonance \cite{kumagaiCeCoIn5}.
An alternative scenario is that $\Delta$ retains a large uniform component, but has an
induced component with modulation ${\bf Q}_i$: $\Delta=\Delta_0 + \Delta_{\mathbf{Q}_i}$, with $2\pi/|{\bf Q}_i| > \xi_0$.
Our data suggest that $\Delta_0$ is large and discontinuous across the first order normal-to-superconducting transition at $T_c\sim 470$ mK at 11.1 T.  The narrow linewidth of the In(1) ($\sim 33$ kHz at 100 mK) suggests that $|\delta K|/(K(T_c)-K(0)) \lesssim 0.07$, where $\delta K$ is the spatial modulation of the Knight shift due to $\Delta_{\mathbf{Q}_i}$.
The absence of any Knight shift modulation at the In(1) preclude any real-space nodes in the order parameter $\Delta$
and are at odds with an exchange-split Fermi surface resulting in an FFLO state.

\section{Discussion}

The NMR data suggest that the B phase of \cecoin\ consists of incommensurate antiferromagnetism coexisting with d-wave
superconductivity in the vortex state.
This material exhibits a large Fermi surface consistent with fully hybridized f electrons \cite{dHvACeCoIn5}, thus the presence of ordered local
moments with magnitude 0.07$\mu_B$
in the superconducting state remains unclear.  Possible explanations include modulated hybridization \cite{PepinNorman} or a two-component scenario \cite{YangDavidPRL}.  The emergence of intrinsic spatial inhomogeneity in pristine undoped materials under these conditions represents a fascinating consequence of proximity to a
first-order quantum critical phase transition,
which may or may not be universal \cite{mackenzieQCP,TusonPNAS}.

We thank R.\ Movshovich, V.\ Mitrovi\'c, A. Balatsky, L. Boulaevskii and M.\ Kenzelmann for valuable discussions and sharing their results.
Work at Los Alamos National Laboratory was performed under the
auspices of the US Department of Energy under grant no.\ DE-AC52-06NA25396.



\end{document}